\documentclass[aps,pre,reprint]{revtex4-1}

\usepackage{amstext}
\usepackage{amsmath}
\usepackage{amssymb}
\usepackage{amsfonts}
\usepackage{units}

\usepackage{graphicx}
\usepackage{color}
\usepackage[normalem]{ulem} 

\renewcommand{\vec}[1]{\boldsymbol{#1}}  
\newcommand*{\diff}{\mathrm{d}}

\newcommand{\zweiervec}[2]{\begin{pmatrix}#1\\#2\end{pmatrix}}
\definecolor{Red}{rgb}{0.9,0.0,0.1}
\definecolor{Blue}{rgb}{0.1,0.0,0.9}
\definecolor{Darkblue}{rgb}{0.22,0.33,0.64}

\begin{document}

\author{Sebastian Knoche} 
\affiliation{Department of Physics, Technische Universit\"{a}t Dortmund, 44221 Dortmund, Germany}

\author{Dominic Vella} 
\affiliation{{Mathematical Institute, Radcliffe Observatory Quarter, Woodstock Road, Oxford, OX2 6GG, UK}}

\author{Elodie Aumaitre} 
\affiliation{Cavendish Laboratory, University of
  Cambridge, JJ Thomson Avenue, Cambridge, CB3 0HE, UK}

\author{Patrick Degen} 
\author{Heinz Rehage} 
\affiliation{Department of
  Chemistry, Technische Universit\"{a}t Dortmund, 44221 Dortmund, Germany}

\author{Pietro Cicuta} 
\affiliation{Cavendish Laboratory, University of
  Cambridge, JJ Thomson Avenue, Cambridge, CB3 0HE, UK}

\author{Jan Kierfeld} 
\email{jan.kierfeld@tu-dortmund.de}
\affiliation{Department of Physics, Technische Universit\"{a}t Dortmund, 44221
  Dortmund, Germany}

\title[Elastometry of Deflated Capsules]{Elastometry of Deflated Capsules: Elastic Moduli from Shape and Wrinkle Analysis}

\begin{abstract}
  Elastic capsules, prepared from droplets or bubbles attached to a capillary (as in a pendant drop tensiometer), can be deflated by suction through the capillary. We study this deflation and show that a combined analysis of the shape and wrinkling characteristics enables us to determine the elastic properties {\it in situ}. Shape contours are analyzed and fitted using shape equations derived from nonlinear membrane-shell theory to give the elastic modulus, Poisson ratio and stress distribution of the membrane. We include wrinkles, which generically form upon deflation, within the shape analysis. Measuring the wavelength of wrinkles and using the calculated stress distribution gives the bending stiffness of the membrane. We compare this method with previous approaches using the Laplace-Young equation and illustrate the method on two very different capsule materials: polymerized octadecyltrichlorosilane (OTS) capsules and hydrophobin (HFBII) coated bubbles.  Our results are in agreement with the available rheological data. For hydrophobin coated bubbles the method reveals an interesting nonlinear behavior consistent with the hydrophobin molecules having a  rigid core surrounded by a softer shell.
\end{abstract}

\maketitle

\section{Introduction}

Elastic capsules are ubiquitous in nature as red blood cells, bacterial or virus capsids, {while} synthetic capsules play an important role in numerous technological applications, including drug delivery and release systems. For stability and applications, the elastic properties of the capsules are crucial, and techniques for the mechanical characterization of single capsules have received much attention (see Refs.~\citenum{Fery2007} {and \citenum{Li2012}} for recent reviews). {Most often, these methods involve contact between the capsule and a probe such as an AFM tip (e.g.\ Refs.\ \citenum{Gordon2004,Arfsten2010,Vella2012}).} However, only very few non-contact techniques are available, and those that are require motion in a surrounding fluid (e.g.\ shape analysis in shear flow \cite{Chang1993} and spinning drop rheometry \cite{Pieper1998}).

Synthetic capsules can be fabricated by various methods \cite{Meier2000}, many of which are based on reactions at interfaces such as polymerization or the adsorption of surfactants \cite{Rehage2002}.
The latter techniques can be applied to enclose a drop or bubble emerging from a capillary within an elastic membrane. A pendant capsule produced in this way can then be deformed by suction through the capillary in order to analyze its elastic response. Because of the simplicity of this procedure, various membrane materials have been studied in this geometry \cite{Husmann2001,Stanimirova2011,Alexandrov2012,Erni2012}. The analysis of those experiments, however, used models developed for pendant drop tensiometry, a technique widely used to determine the surface tension of  {\em liquid-liquid interfaces} by fitting the drop shape to that predicted by the Laplace-Young equation \cite{Rotenberg1983}. This technique is not valid for elastic capsules, since it neglects the elastic stresses within the membrane{\cite{Carvajal2011,Ferri2012}.}

In this article, we present a non-contact elastometry method for individual capsules, inspired by the pendant drop method but adjusting the theoretical model to account for elasticity. Elastic capsules that are attached to a capillary are deflated by sucking some of the enclosed medium back into the capillary. We describe deflated shapes (see Fig.\ \ref{fig:plot1} for examples) using shell theory for axisymmetric membranes and accounting for the wrinkling induced by deflation. By analyzing the capsule's shape and wrinkling pattern, we can determine its elastic properties, namely the surface Young modulus $Y_{2D}$, describing the membrane's resistance to stretching, and the Poisson ratio $\nu_{2D}$ describing the lateral contraction upon stretching. By adjusting these parameters, the theoretical capsule contour can be fitted to the observed shape (Figs.\ \ref{fig:plot1}(a) and (b)), and the elastic moduli of the membrane can be determined over a whole range of capsule volumes. 

The determination of the  membrane's bending stiffness $E_B$ represents another challenge for elastic capsules because it only has a small influence on the capsule contour on large scales and, thus, cannot be obtained by  the fitting. Therefore, we combine the shape analysis described above with an analysis of the wrinkles that generically form during deflation, and deduce the bending stiffness from the wavelength of the wrinkles \cite{Cerda2003}. This combined approach enables us to determine all elastic constants of individual capsules {\it in situ} from images of the initial and deflated capsule, thus offering a valuable alternative to rheology in planar geometries. In particular, the capsules studied here have a geometry similar to that of capsules used in applications in pharmacy or industry.

\section{Materials and Methods}
\subsection{Elastic {M}odel}
\begin{figure}[t]
  \centerline{\includegraphics[width=82.5mm]{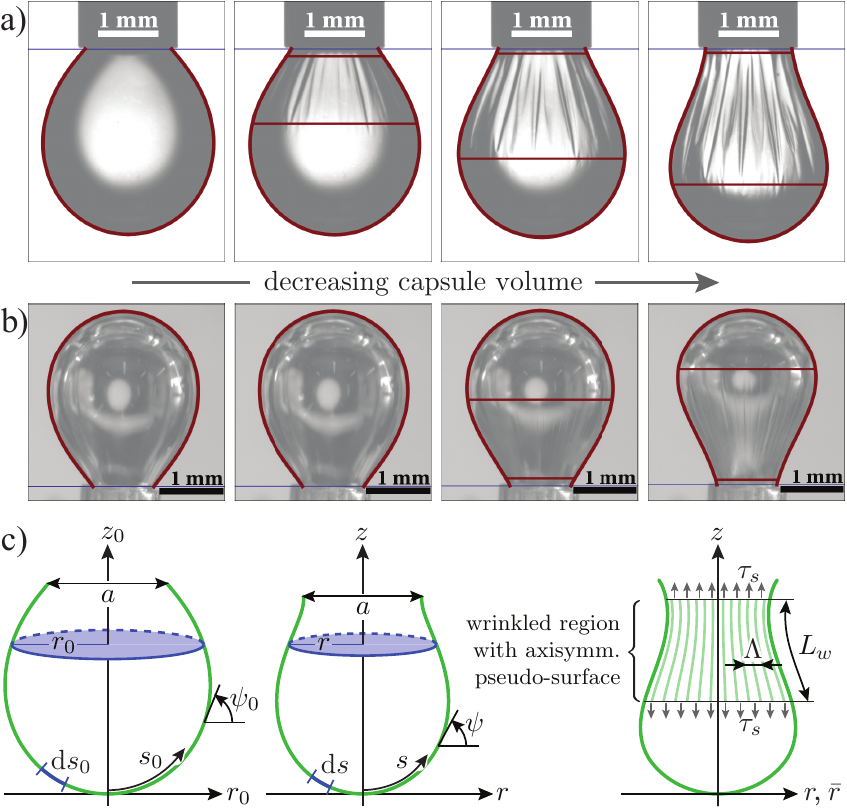}}
  \caption{(a) and (b) Comparison between fitted contour (solid curve) and original image for OTS (a) and HFBII (b) capsules. The first image shows the equilibrated capsule, while the next three show increasing levels of deflation. The extent of the wrinkled region predicted by the model is shown by the horizontal lines. (c) Arc-length parametrizations in cylindrical coordinates of the undeformed $(r_0(s_0), z_0(s_0))$, deformed $(r(s), z(s))$ and wrinkled midsurface (from left to right).
  }
\label{fig:plot1}
\end{figure}

We model the capsule as an elastic membrane covering a droplet or bubble, which is attached to a capillary of diameter $a$. We neglect the bending resistance for simplicity. The axisymmetric reference configuration (Fig.\ \ref{fig:plot1}(c), left) is assumed to be free of elastic stresses; the capsule shape is determined by the balance between an isotropic interfacial tension $\gamma$ and gravity, which is described by the Laplace-Young equation \cite{Landau1987}. We make this assumption because the elastic capsule is formed in this initial state from a fluid interface. {In the undeformed state, gravity causes the capsule to form the tear drop shape seen in fig.\ \ref{fig:plot1}. In the absence of gravity, it is well-known that the capsule would take the shape of a spherical cap}{; we therefore incorporate gravity in our analysis.}

Upon deflation, the capsule changes to a deformed configuration, {which we assume is axisymmetric} (Fig.\ \ref{fig:plot1}(c), middle). The local deformation is measured by the meridional and hoop stretches given by
\begin{equation} \label{eq:strains}
 \lambda_s =\diff s / \diff s_0 \quad \text{and} \quad \lambda_\phi = r/r_0
\end{equation} 
respectively (see Fig.\ \ref{fig:plot1}(c)). They lead to elastic tensions according to a Hookean constitutive relation \cite{Libai1998, Knoche2011}, which reads for the meridional tension
\begin{equation} \label{eq:hooke}
 \tau_s = \frac{Y_{2D}}{1-\nu_{2D}^2} \, \frac{1}{\lambda_\phi} 
    \big[ (\lambda_s-1) + \nu_{2D}\, (\lambda_\phi-1) \big] + \gamma 
\end{equation}
with the surface Young modulus $Y_{2D}$ (which for isotropic media is related to the bulk Young modulus $Y_{3D}$ and membrane thickness $H_0$ by $Y_{2D}=Y_{3D}H_0$) and surface Poisson ratio $\nu_{2D}$. The surface (two-dimensional) Poisson ratio is, for stability reasons, limited to the range $-1 < \nu_{2D} < 1$, as opposed to the bulk (three-dimensional) Poisson ratio which is confined to the range $-1 < \nu_{3D} < 1/2$ \cite{Barthes-Biesel2002,Landau1987}. The constitutive law for the hoop tension $\tau_\phi$ is obtained by interchanging all indices $s$ and $\phi$. An equilibrium capsule configuration has to satisfy the force balance equations
\cite{Libai1998}
\begin{equation} \label{eq:force_balance}
 \begin{aligned}
 0 &= - \frac{\cos \psi}{r}\, \tau_\phi 
  + \frac{1}{r}\, \frac{\diff(r\,\tau_s)}{\diff s} \\
 p-\Delta\rho \, g \, z &=  \kappa_\phi \, \tau_\phi 
   + \kappa_s \, \tau_s.
 \end{aligned}
\end{equation} 
Here, $p$ is the pressure {inside} the capsule {at the} apex, and $\Delta\rho \, g \, z$ {is} the pressure contribution caused by {gravity and} the density difference between the inner and outer fluids.
The principal curvatures are denoted by $\kappa_s$ and $\kappa_\phi$, and the slope angle $\psi$ is defined in Fig.\ \ref{fig:plot1}(c). This system of differential equations must be solved numerically subject to boundary condition{s that fix} the capsule radius to the inner radius of the capillary, $a/2${, and ensure that the capsule is closed and smooth at its apex.} {Using these boundary conditions, eq.\ (\ref{eq:force_balance}) determines the capsule shape (and, therefore, also the capsule volume $V$) for given material parameters $Y_{2D}$ and $\nu_{2D}$ and given pressure $p$.} More details are given in Appendix I.

If the capsule is deflated sufficiently, regions with compressive hoop stress (i.e., $\tau_\phi < 0$) develop and wrinkles form in order to release this stress, which a membrane with small bending modulus cannot support \cite{Davidovitch2011,Vella2011,Basheva2011,King2012}. For fully developed wrinkles the hoop stress is almost completely relaxed \cite{Davidovitch2011}, and so we modify the shape equations by setting $\tau_\phi = 0$ in the wrinkled region. Assuming wrinkles of small amplitude, the membrane can be described by an axisymmetric pseudo-surface with radial coordinate $\bar r(s)$ around which the wrinkled non-axisymmetric midsurface {oscillates} \cite{Libai1998}, see also the Appendix I. Using the condition $\tau_\phi=0$ in eq.\ (\ref{eq:hooke}) to find $\lambda_\phi(\lambda_s)$ we then obtain a modified expression for $\tau_s$ in the wrinkled region in terms of $\lambda_s$, in which $\lambda_\phi$ is eliminated. This allows us to obtain a closed set of modified shape equations in the wrinkled region by applying the axisymmetric force balance (\ref{eq:force_balance}) to the pseudo-surface. Theoretical axisymmetric shapes obtained from integrating eqs.\ (\ref{eq:force_balance}) can then be fitted to experimental images by varying the two material parameters $Y_{2D}$ and $\nu_{2D}$ and the pressure $p$.

We have not included a bending energy in {the model described above}, because for thin membranes with a small bending modulus ($E_B\propto Y_{2D} H_0^2$ in the case of isotropic materials) the bending moments give only small corrections in the shape equations (\ref{eq:force_balance}). These corrections are controlled by the dimensionless parameter $E_B/\gamma a^2$. Using $a \sim 1 \, \rm mm$ as the capillary diameter, we find that this parameter is only of the order of $10^{-6}$ for OTS and $10^{-10}$ for HFBII capsules. Therefore, $E_B$ cannot be inferred directly from an analysis of the capsule's {shape}.

\subsection{Wrinkle {W}avelength}
The shape equations can predict the regions where wrinkles will occur, but not their amplitude and wavelength. These characteristics are mainly determined by the bending modulus $E_B$ of the membrane. 

\begin{figure}[t]
  \centerline{\includegraphics[width=82.5mm]{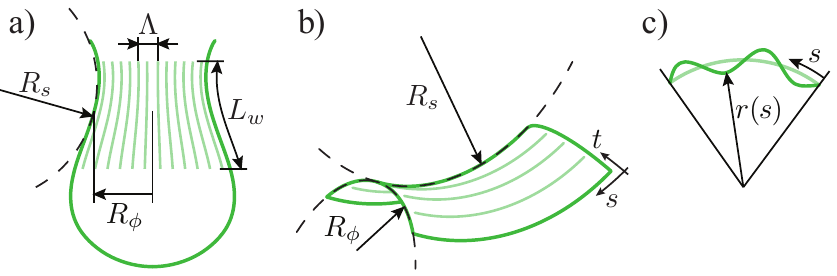}}
  \caption{a) The wrinkled region of a capsule is curved in the meridional and circumferential directions. b) Geometry for the analytic calculation of the deformation energies. The membrane patch has two radii of curvature, $R_s$ and $R_\phi$ and is parametrized via the arc lengths $s$ (in meridional direction) and $t=\phi R_\phi$ (in circumferential direction). c) An initially curved fibre (in either $s$ or $t$ direction) is wrinkled by adding a sinusoidal normal displacement.}
\label{fig:wrinkling}
\end{figure}

As shown in Fig.\ \ref{fig:wrinkling}, the wrinkled region is curved in both meridional and circumferential direction with curvatures $\kappa_s = 1/R_s$ and $\kappa_\phi = 1/R_\phi$, which we assume to be approximately constant. Within this region, we assume a homogeneous state of stress with tensional $\tau_s>0$ and compressive $\tau_\phi < 0$. The wavelength of the wrinkles can be determined by balancing the main contributions to the deformation energy: bending in the circumferential direction and stretching in both circumferential and meridional directions. {Changes in the gravitational potential energy caused by wrinkling are neglected since the wrinkles are largely parallel to the $z$-axis.}

Upon wrinkling, the membrane is displaced sinusoidally in the normal direction resulting in local strains $\varepsilon_s=|\partial \vec r/\partial s|-1$ and $\varepsilon_\phi=|\partial \vec r/\partial t|-1$ (see Fig.\ \ref{fig:wrinkling}). As they are working against the meridional and circumferential tension, respectively, the stretching energy during the formation of wrinkles is
\begin{equation}
 W_S = \int \diff s \, \diff t \left\{\tau_s \varepsilon_s + \tau_\phi \varepsilon_\phi \right\}.
\end{equation}
The bending energy is mainly determined by the curvature change $\Delta \kappa_\phi$ in the circumferential direction and reads
\begin{equation}
 W_B = \int \diff s \, \diff t \left\{ \frac{1}{2} E_B (\Delta\kappa_\phi)^2 \right\}.
\end{equation}
The evaluation of these integrals is performed in Appendix II. {Wrinkling occurs because of a competition between the increase in elastic energy caused by bending and meridional stretching and the decrease in energy achieved by releasing the compressive stress $\tau_\phi$.} The wrinkled state becomes energetically preferable if $W_S + W_B < 0$, corresponding to (see Appendix II)
\begin{equation}
 \tau_\phi < -\tau_s \frac{\Lambda^2}{4 L_w^2} - E_B \frac{\Lambda^2}{4 \pi^2}\left( \frac{4 \pi^2}{\Lambda^2} - \frac{1}{R_\phi^2} \right)^2,
 \label{eq:tau_phi}
\end{equation}
where $L_w$ is the length of the wrinkles and $\Lambda$ their wavelength, see Fig.\ \ref{fig:wrinkling} (a).

The most unstable wrinkling mode has a wavelength $\Lambda_c$ that minimizes $|\tau_\phi(\Lambda)|$,
\begin{align}
 \Lambda_c &= \left( \frac{16 \pi^2 E_B L_w^2}{\tau_s + E_B L_w^2 / \pi^2 R_\phi^4} \right)^{1/4} \nonumber \\
 &\approx \left( \frac{16 \pi^2 E_B L_w^2}{\tau_s} \right)^{1/4}.
 \label{eq:wavelength}
\end{align} 
The {final} approximation holds if $ R_\phi \gg \Lambda$, which is clearly the case in the experiments presented here (see Fig.\ \ref{fig:plot1}). The approximated form agrees with the results of Ref.\ \citenum{Cerda2003} for a planar geometry.

\subsection{Fitting {P}rocedure}
Based on the theory presented above, a three step fitting procedure can be used in order to determine the elastic moduli of the capsule membrane:
\begin{enumerate}
  \item The undeformed  capsule shape is fitted using the Laplace-Young equation with  the interfacial tension $\gamma$ and pressure $p_0$ inside the capsule as free parameters.

  \item Shape analysis: Solutions of the shape equations (\ref{eq:force_balance}) are fitted to images of the capsule with $p$, $\nu_{2D}$ and $K_{2D} = Y_{2D}/ 2(1-\nu_{2D})$ (area compression modulus) as free parameters {at each stage of deflation}.

  \item Wrinkle analysis: The wavelength $\Lambda$ in the center of the wrinkled region is measured from images. The length $L_w$ of the wrinkles and a mean value of  $\tau_s$ over this region are obtained from the fitted solution. Then, the bending modulus $E_B$ is determined from eq.\ (\ref{eq:wavelength}) (or (\ref{eq:EB}), see Appendix II). Using the relationship $E_B = Y_{2D} H_0^2 / 12(1-\nu_{3D}^2)$ from classical shell theory \cite{Landau1986} an effective membrane thickness $H_0$ can also be estimated \footnote{This is a rough estimate, since this relation applies only to thin sheets of isotropic material, but the membranes at hand are evidently anisotropic.}.
\end{enumerate}
Note that the position and  height $L_w$ of the wrinkled region are {\em not} fit parameters but can be used as an independent check {of the goodness of the fit}. Technical details of the fitting procedure and the underlying image analysis are contained in the SI. {$L_w$ is determined from the fitted numerical solution as the arc length over which the modified shape equations (with $\tau_{\phi} = 0$) were integrated}{, see the Appendix I.}

We {now} demonstrate this method on two rather different types of capsules: polymerized OTS capsules and bubbles coated with an interfacial monolayer film of the protein hydrophobin.

\subsection{Preparation of {C}apsules}
To prepare a pendant capsule, a glass cell is filled with p-xylene containing OTS. Then a drop of water is placed into this phase using a syringe. The polymerization process starts immediately after the oil/water-interface is formed. Hydrophobin coated bubbles are prepared in a very similar fashion. As described in previous work \cite{Aumaitre2013}, an air bubble is placed into a solution of HFBII in water using a J-shaped needle {and HFBII molecules adsorb at the interface over the course of 20 minutes}.

After equilibration, the capsules are deflated slowly (i.e. quasi-statically{, on a timescale of $\sim 10 \, \rm s$ for a deflation of OTS capsules and even slower for HFBII capsules}) by sucking the enclosed medium back into the syringe. The OTS capsule is subsequently re-inflated to check whether the deformation is reversible.

\section{Experiments and Results}

\subsection{Test of Shape Analysis}
We first test our elastometry approach by applying it to fit numerically generated capsule shapes. To this end we take an initial capsule configuration ($\gamma = 49.8\,\rm mN/m$, $\Delta\rho_0 = 1000\,\rm kg/m^3$ and $V_0=8.23 \, \rm mm^3${, values taken from the HFBII capsule}) and use our shape equations with fixed elastic moduli ($K_{2D} = 600\, \rm mN/m$, $\nu_{2D}=0.3$) to compute deflated configurations. From a contour, we calculate a set of approximately 150 sampling points, optionally add some noise to simulate an imperfect contour analysis, and pass them to the fitting procedure to see if it finds the correct solution.

\begin{figure}[t]
  \centerline{\includegraphics[width=82.5mm]{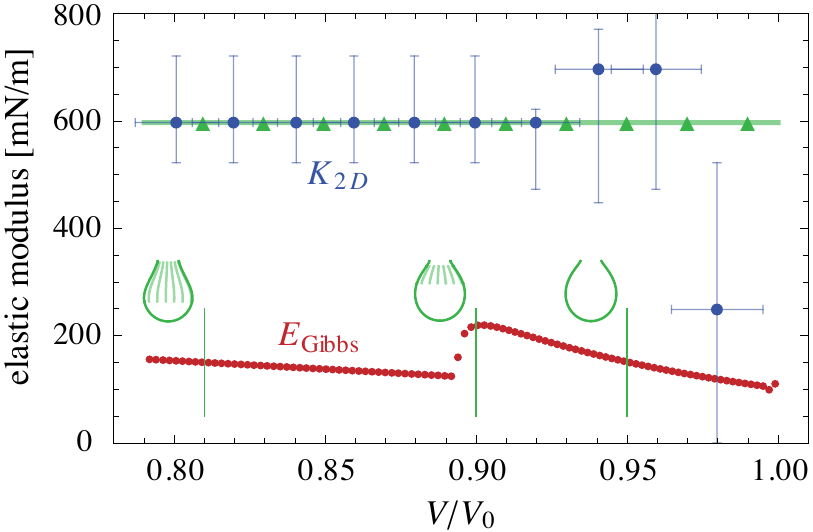}}
  \caption{Fit results for theoretically generated capsule shapes with $K_{2D} = 600 \, \rm mN/m$ (green line). Green triangles represent the fits of the elastic shape equations to the clean contour, blue dots the elastic fits to the noisy contour, and red points the Gibbs elasticity calculated from Laplace-Young fits. Vertical green lines indicate the positions of the capsule pictograms shown in green.}
\label{fig:Theorie_Fits}
\end{figure}

Fig.\ \ref{fig:Theorie_Fits} shows that all fits to the clean contour (green triangles) are successful and recover the original compression modulus.
The elastic fits to the noisy contour (blue points) succeed if the deformation is large enough, i.e.\ for $V/V_0 \le 0.94$ in the present case. For smaller deformation, there are some deviations in the fit results, but the error bars are large enough to reach the real value, except for the very first fit ($V/V_0=0.98$). This problem arises due to the very small deformation: The root mean square deviation between the initial shape and the shape at $V/V_0=0.98$ is about 0.01 length units, the noise amplitude is 0.005 and the offset used for the error bars is $\pm$ 0.007 (corresponding to $\pm$ 1 pixel at usual image resolution). So the sampling points passed to the fitting procedure have an offset from their original place which is of the same order as the deformation; we could not have expected the fits to work.

\begin{figure*}[t]
  \includegraphics[width=177.8mm]{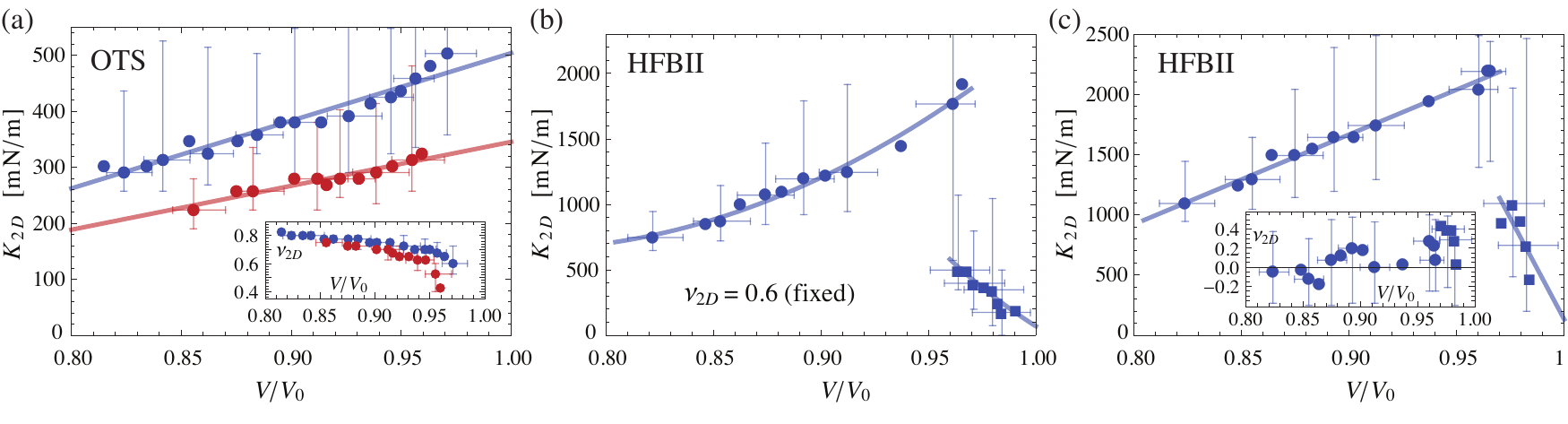} 
  \caption{
   Fit results for (a) an OTS capsule and (b, c) an HFBII capsule, with non-wrinkled ($\scriptscriptstyle\blacksquare$) and wrinkled ($\bullet$) shapes. Error bars were generated by displacing the sampling points about $\pm 1$ pixel, see SI. Lines are drawn to guide the eye. (a) Upper blue curve is for deflation, lower red curve for re-inflation. {The deflation was driven to even smaller volumes than shown, but for these the contour analysis failed.} (b) For HFBII, the Poisson ratio was fixed to $\nu_{2D} = 0.6$. (c) The same HFBII capsule fitted with free Poisson ratio.
  }
 \label{fig:fitresults}
\end{figure*}

\subsection{Comparison to Laplace-Young Analysis}

For comparison with our new method, we also consider the performance of the usual Laplace-Young analysis used by many scientists \cite{Husmann2001,Stanimirova2011,Alexandrov2012}. In a Laplace-Young analysis, elastic capsules are fitted with the Laplace-Young equation to obtain the interfacial tension $\gamma$ and capsule surface $A$ over the course of the deflation; these tools are provided by the software of common pendant drop tensiometers. Usually, the Gibbs elastic modulus $E_\text{Gibbs} = A \, \diff\gamma/\diff A$ is then calculated from these values. In applying this method to theoretically generated capsules shapes, we find that $E_\text{Gibbs}$ is significantly smaller than the actual area compression modulus, see Fig.\ \ref{fig:Theorie_Fits}. It appears that for elastic capsules, the intricate interplay between membrane geometry and elastic tensions renders the Laplace-Young analysis more erroneous than intuitively expected: Not even for small deformation does the Gibbs elastic modulus approach the real area compression modulus. This explains the observations of Stanimirova et.\ al.\ that pendant drop tensiometry gives wrong results if applied to capsules with high surface elasticity \cite{Stanimirova2011}.

\subsection{Fit results for OTS and HFBII capsules}

For OTS and HFBII capsules, several images of the undeformed reference configurations are fitted to the solution of the Laplace-Young equation. The results are averaged to obtain the surface tension $\gamma = 11.2\,{\rm mN/m}$ for OTS and $\gamma = 49.8\,{\rm mN/m}$ for HFBII. Both values are lower than the respective values of the clean interfaces {because the OTS and HFBII molecules are surface active agents that lower the interfacial tension during adsorption}.

Deflated capsule configurations with varying volume are fitted using the elastic model. Fig.\ \ref{fig:fitresults}(a) shows the results for an OTS capsule. All data points in Fig.\ \ref{fig:fitresults}(a) represent wrinkled shapes, because even the slightest deformation gives rise to wrinkles due to the low initial surface tension and high compression modulus. We find an  area compression modulus $K_{2D}$ which decreases with decreasing $V/V_0$. Although the error bars in Fig.\ \ref{fig:fitresults}(a) are overlapping, this result is reliable because the error bars represent worst case systematic errors (see SI). The deformation is not perfectly reversible, and we observe hysteresis: The area compression modulus obtained for re-inflated capsules is lower (lower red vs.\ upper blue data points in Fig.\ \ref{fig:fitresults}(a)). The presence of hysteresis indicates that the decreasing modulus is not an artifact of the method but a result of creep, for example  by viscous effects, i.e.\ breakage or rearrangement of bonds in the OTS network, or by the formation of micro-defects such as shear cracks. The video in the SI shows, however, that computed contours with the moduli fixed to the small-deformation values $K_{2D} \approx 500\,{\rm mN/m}$ and $\nu_{2D} \approx 0.6$ are in good agreement with all experimental observations, implying that the nonlinear effects are moderate. The resulting surface shear modulus \cite{Barthes-Biesel2002} is $G_{2D} = K_{2D} (1-\nu_{2D})/(1+\nu_{2D}) \approx 125\, \rm mN/m$. In Ref.~\citenum{Rehage2002}, larger values of $200 - 300\, \rm mN/m$ (obtained by interfacial shear rheology) are reported for similar OTS membranes. {In another experiment with 3 deflation/inflation cycles of an OTS capsule we saw that the capsule does not weaken further after the first deflation, but hysteresis was observed in all cycles. The hysteresis may possibly depend on the deflation velocity and may thus contain information about the viscous part of the membrane visco-elasticity, this issue is left for future research.} {Since all viscous effects have been neglected in the elastic model, our analysis should only be applied to quasi-static experiments.}

In the case of HFBII, we can reduce the number of fit parameters by constraining $\nu_{2D}=0.6$ to a value measured in an independent experiment \cite{Aumaitre2012thesis} and determine the area compression modulus only. Fig.\ \ref{fig:fitresults}(b) shows that the area compression modulus $K_{2D}$ increases for small deformations, where the capsule does not wrinkle (blue squares in Fig.\ \ref{fig:fitresults}(b)), to values around $500\,{\rm mN/m}$. The onset of wrinkling coincides with a sharp increase of the modulus to a maximum value of $K_{2D} \approx 2000\,{\rm mN/m}$. This sharp increase is consistent with the molecular structure of HFBII \cite{Hakanpaa2004}, which contains a rigid core consisting of four $\beta$-strands and is stabilized by disulfide bridges. The modulus $K_{2D}$ increases sharply when compression of this rigid protein core sets in, while at small deformations, only contacts between hydrophobin proteins or a soft shell consisting of coil and loop structures surrounding the rigid $\beta$-barrel are compressed. The sharp rise of the compression modulus triggers wrinkling. Subsequently, the compression modulus decreases again (blue circles in Fig.\ \ref{fig:fitresults}(b)) likely signalling creep as also observed for the OTS capsules. Possible explanations for the creep behavior are the formation of micro-defects such as shear cracks or localized bulges into the subphase, which weaken the hydrophobin layer.

The choice of the fixed value for $\nu_{2D}$ influences the absolute values obtained for $K_{2D}$ and the size of its jump when wrinkling sets in, while the characteristics described above are robust. Taking the Poisson ratio as a fit parameter also results in a similar course of the elastic modulus, see Fig.\ \ref{fig:fitresults}(c). And yet the results for $\nu_{2D}$ differ substantially from the previously assumed value of $\nu_{2D}=0.6$. Especially for small deformations, this results in higher values for the area compression modulus.

The values $K_{2D} < 500\,{\rm mN/m}$ for the compression modulus for small deformations and prior to wrinkling are in good agreement with values reported previously for HFBII \cite{Cox2007,Blijdenstein2010,Alexandrov2012}. The large values around $K_{2D} = 2000\,{\rm mN/m}$ at the onset of wrinkling have not been reported before, since the experimental methods used in the literature are not reliable in the presence of wrinkles. However, a comparison to viral capsids consisting of densely packed proteins is possible. In Ref.~\citenum{Ivanovska2004}, the bulk Young modulus of a viral capsid is measured as $1.8 \, \rm GPa$, which is comparable to our result for the bulk modulus $Y_{3D} = Y_{2D} / H_0 \approx 1\, \rm GPa$, where $H_0 \approx 2\,\rm nm$ is the hydrophobin layer thickness\cite{Kisko2009}.

\subsection{Analysis of the {W}rinkle {W}avelength}

\begin{figure}[t]
  \centerline{\includegraphics[width=82.5mm]{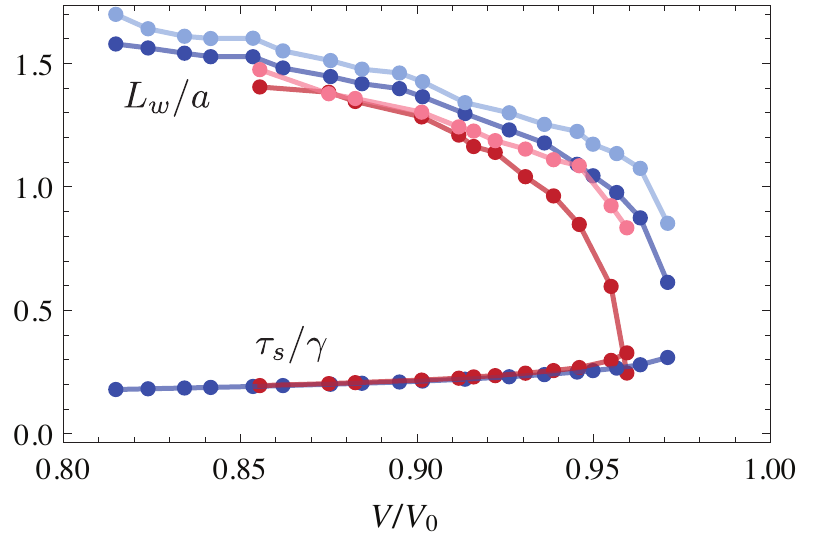}}
  \caption{Wrinkle length $L_w$ (upper curves) and mean tension $\tau_s$ (lower curves), both in reduced units for the wrinkled OTS capsule; the unit length is $a = 1.44 \, \rm mm$ and tension unit $\gamma = 11.2 \, \rm mN/m$. Blue dots indicate the deflation results, and red dots the re-inflation. Dark blue and dark red represent results taken from the fitted shape equations, light blue and light red represent direct measurements from the images.}
\label{fig:wrinkle_Lw_tau}
\end{figure}

Finally, the wrinkling pattern shall be analyzed and be related to the bending stiffness. The wavelength of the wrinkles on the HFBII capsules cannot be measured directly: due to the low bending stiffness, the wavelength is too small to be resolved in the {experimental} images {(note that the observable folds in Fig.\ \ref{fig:plot1}(b) are not primary wrinkles, but rather secondary or higher order structures)}. However, using eq.\ (\ref{eq:wavelength}) with $L_w$, $\tau_s$ and $E_B = Y_{2D} H_0^2 / 12(1-\nu_{3D}^2)$ obtained from the elastic fits, we expect wavelengths between $7.6$ and $11 \, \rm \mu m$. In the literature, similarly small or even smaller wrinkle wavelengths for compressed HFBII films in a Langmuir trough have been reported \cite{Basheva2011,Blijdenstein2010,Aumaitre2012thesis}.

For OTS capsules, however, wrinkle wavelengths $\Lambda$ may be determined from images. Values for $\tau_s$ in the wrinkled region and $L_w$ are taken from the elastic fits and are documented in Fig.\ \ref{fig:wrinkle_Lw_tau} (dark blue and dark red dots). The resulting bending stiffness is $E_B \approx (2.5 \pm 0.7) \cdot 10^{-14}\,{\rm Nm}$; {that is three orders of magnitude larger than previous estimates\cite{Cerda2003,Finken2006} which used an experiment with a smaller capsule of the same material in shear flow, resulting in wrinkles with a shorter wavelength\cite{Walter2001}.} {Moreover, OTS capsules in Ref.\ \citenum{Walter2001} were prepared in a different aqueous solution (glycerol and NaOH) and using longer polymerization times. These differences can give rise to distinct membrane thicknesses and crosslink densities, which can explain the differences in the bending modulus: the bending modulus varies with the third power of the membrane thickness and Cerda and Mahadevan estimated in Ref.\ \citenum{Cerda2003} a thickness around 20 times smaller than that of the capsules used here.} Combining this value for $E_B$ with measurements of $Y_{2D}$ from the shape analysis we estimate the membrane thickness $H_0 \approx (0.77 \pm 0.07) \,{\rm \mu m}$, which is in approximate agreement with capsule thicknesses $0.86<H_0 <1.4\,{\rm \mu m}$ measured by electron microscopy, see the SI. Also the extent and position of the wrinkled region are in good agreement with the experimental data, see Fig.\ \ref{fig:wrinkle_Lw_tau}: Except for the last part of the re-inflation curve, the curves for the wrinkle length of the elastic fits (dark symbols) and for the directly measured wrinkle length (light symbols) are quite close to each other.

\section{Conclusions}
The proposed theory for axisymmetric capsule shapes in the presence of wrinkling describes deflated experimental shapes of both OTS capsules and hydrophobin coated bubbles accurately. It is possible to fit the solutions of the  shape equations to contours extracted from experimental images in order to find the elastic properties (area compression modulus and Poisson ratio) of the membrane. Additionally, a subsequent analysis of the wavelength of wrinkles which occur during the deflation can determine the membrane's bending stiffness. With this combination of analyses, the elastic properties of the capsule are completely characterized.

Applying this method to OTS capsules gives reasonable values for all three elastic constants{:} $K_{2D} \approx 500\,{\rm mN/m}$, $\nu_{2D} \approx 0.6$ and $E_B \approx 2.5 \cdot 10^{-14}\,{\rm Nm}$ for the small deformation behavior. Furthermore, we observe a softening or creep of the capsules with decreasing volume, which we also observe for hydrophobin capsules (see Fig.\ \ref{fig:fitresults}). 

For hydrophobin capsules, the area compression modulus initially grows upon deflation, $K_{2D} = 160\,{\rm mN/m} \text{ to } 500\,{\rm mN/m}$ when we assume $\nu_{2D} = 0.6$. At the onset of wrinkling, it jumps to $2000\,{\rm mN/m}$ because compression of the rigid protein core sets in (see Fig.\ \ref{fig:fitresults} (b) and (c)). Obviously, this complex behavior cannot be explained by simple Hookean elasticity, and we hope that these results will inspire future work on possible nonlinear elastic laws for HFBII membranes or other membrane materials consisting of hard core particles. According to our observations, this should include an immense strain stiffening upon compression. 
In the application of the Laplace-Young analysis to elastic capsules, we found that the shape analysis reacts delicately to inaccuracies in the model for the membrane tensions. Likewise, a certain amount of caution is advisable when our elastometry method indicates a strongly non-linear elasticity, as in the present analysis of the HFBII capsule. In this case, obviously non-linear elasticity is fitted with a simple Hookean constitutive law, and we cannot be sure which characteristics of the results reflect limitations of the {linearly} elastic model.

These two  applications prove the concept of the elastometry method, which could be added to the features of pendant drop tensiometers in the future. The method can reveal changes in elastic constants  with decreasing volume that are not accessible by other methods. It can be further improved by using data from a simultaneous pressure measurement during deflation, which would eliminate one of the fit parameters.

\begin{acknowledgments}
SK and JK  acknowledge financial support by the Mercator Research Center Ruhr (MERCUR). DV is partially supported by  a Leverhulme Trust Research Fellowship. EA and PC thank Unilever Global Development Centre for the gift of hydrophobin and EPSRC and Unilever, plc for funding.  
\end{acknowledgments}

\section*{Supporting Information}
 Technical details on the image analysis and fitting procedure are contained in the supporting information. {A video of the OTS experiment and fitted shape equations (green line, with $K_{2D} = 500 \, \rm mN/m$ and $\nu_{2D}=0.6$ fixed) is also contained; the blue lines indicate the computed amplitude of the wrinkles.}

\appendix

\section{Shape Equations}
\subsection{The Non-Wrinkled Case}
In this appendix, we show how the elastic model, defined in the main text by Fig.\ \ref{fig:plot1} (c) and equations (\ref{eq:strains}), (\ref{eq:hooke}) and (\ref{eq:force_balance}), can be treated numerically. This is best handled when a system of first order differential equations is constructed from the force balance, constitutive and geometrical equations.

The axisymmetric reference configuration of the pendant or rising capsule is described in cylindrical coordinates by a midsurface parametrisation $(r_0(s_0), z_0(s_0))$ with $s_0 \in [0, L_0]$ being the arc length. It is free of elastic tensions; the capsule retains its shape only because of an isotropic interfacial tension $\gamma$. Accordingly, the reference shape is described by the Laplace-Young equation \cite{Landau1987}
\begin{equation}
 \gamma (\kappa_{s0} + \kappa_{\phi0}) = p_0 - \Delta \rho g z_0
\end{equation} 
where $\kappa_{s0}$ and $\kappa_{\phi0}$ are the principal curvatures and $p_0 - \Delta \rho g z_0$ is the hydrostatic pressure caused by the density difference of inner and outer fluid.

When exerting forces on the capsule, it changes to a deformed configuration $(r(s_0), z(s_0))$, with the so called ``material coordinate'' $s_0$ still running from $0$ to $L_0$. Alternatively, we can choose an arc length parametrisation $(r(s), z(s))$ plus a mapping $s(s_0)$ to describe this configuration.

In the latter notation, some geometric relations can be written quite conveniently. For later use we introduce the slope angle $\psi$ (see Fig. \ref{fig:plot1} c) defined by
\begin{equation} \label{eq:slope_angle}
 \cos \psi = \diff r / \diff s \quad \text{and} \quad \sin \psi = \diff z / \diff s,
\end{equation} 
and the principal curvatures 
\begin{equation} \label{eq:curvatures}
 \kappa_s = \diff \psi / \diff s \quad \text{and} \quad \kappa_\phi = \sin \psi / r.
\end{equation} 
Together with the force balance (\ref{eq:force_balance}), strain definition (\ref{eq:strains}) and elastic law (\ref{eq:hooke}), these relations can be used to construct a system of first order differential equations with the material coordinate $s_0$ as variable,
\begin{equation}
 \begin{aligned}
  r'(s_0) &= \lambda_s \, \cos \psi \\
  z'(s_0) &= \lambda_s \sin \psi \\
  \psi'(s_0) &= \frac{\lambda_s}{\tau_s} \left( p - \Delta \rho \, g \, z - \kappa_\phi \, \tau_\phi \right) \\
  \tau_s'(s_0) &= -\lambda_s \, \cos \psi\,\frac{\tau_s - \tau_\phi}{r} . \label{eq:shape_eqns}
 \end{aligned}
\end{equation} 
All functions occurring on the right hand side of the system must be expressed in terms of the basic functions $r$, $z$, $\psi$, $\tau_s$ via the previously mentioned geometric relations and definitions of stretches and tensions. The boundary conditions are obvious from the geometry of the capsule, $r(0) = z(0) = \psi(0) = 0$ and $r(L_0) = a/2$, where $a$ is the inner diameter of the capillary. Finally, some limits for $s_0 \rightarrow 0$ must be evaluated analytically using L'H\^ospital's rule to start the integration. For the nondimensionalization, we choose the capillary diameter $a$ as the length unit and the interfacial tension $\gamma$ of the initial shape as tension unit.

\subsection{Extending the Model to Wrinkled Shapes}
Now we want to calculate wrinkled configurations (with wave vector along the hoop direction). Configurations of this kind arise because ideal membranes without bending resistance cannot support negative tensions. They are not exactly axisymmetric, but can be approximated by an axisymmetric pseudo-surface in the wrinkled region. The shape $(\bar r(s_0), \bar z(s_0))$ of the pseudo-surface is obtained by requiring $\tau_\phi = 0$ in areas where the original model would yield $\tau_\phi < 0$ \cite{Libai1998}.

According to Hooke's law (\ref{eq:hooke}), the wrinkling condition $\tau_\phi < 0$ is equivalent to
\begin{equation}
\lambda_\phi < 1- \gamma \frac{1-\nu_{2D}^2}{Y_{2D}} \lambda_s - \nu_{2D} (\lambda_s - 1). \label{eq:wrinkling_cond}
\end{equation}
At the point where $\lambda_\phi$ falls below this threshold during the numeric integration of (\ref{eq:shape_eqns}), we switch to a modified system of shape equations to continue the integration. This system describes the pseudo-surface and is mainly determined by setting $ \tau_\phi = 0 $ on the wrinkling domain, i.e.
\begin{equation} \label{eq:lambda_phi_lambda_s}
 \lambda_\phi =  1 - \gamma \frac{1-\nu_{2D}^2}{Y_{2D}} \lambda_s - \nu_{2D} (\lambda_s - 1).
\end{equation}
Note that $\lambda_\phi$ is the hoop stretch of the real, wrinkled midsurface and not to be confused with the stretch $\bar \lambda_\phi = \bar r / r_0$ of the pseudo-surface (all quantities referring to the pseudo-surface are indicated with an overbar).

In order to eliminate the hoop stretch of the real midsurface from our system of equations, we insert this expression (\ref{eq:lambda_phi_lambda_s}) into the constitutive relation (\ref{eq:hooke}) for the meridional tension,
\begin{equation}
 \tau_s = Y_{2D} \frac{1}{\lambda_\phi} \left[ (\lambda_s-1) - \frac{\nu_{2D}\gamma}{Y_{2D}} \lambda_s  \right] + \gamma.
\end{equation}
However, this tension is not suitable for considering the force balance since it is measured per unit length of the wrinkled, non-axisymmetric midsurface. In order to adopt the simple force balance (\ref{eq:force_balance}) for the pseudo-surface, we have to measure the tension per unit length of the pseudo-surface, $\bar \tau_s = \tau_s {\lambda_\phi}/{\bar \lambda_\phi}$, resulting in
\begin{multline}
 \bar \tau_s = \frac{1}{\bar \lambda_\phi} \Big[ \lambda_s \Big( Y_{2D} -2 \nu_{2D} \gamma-\gamma^2\frac{1-\nu_{2D}^2}{Y_{2D}}\Big) \\ - Y_{2D} + \gamma (1+\nu_{2D}) \Big].
\end{multline} 

With this constitutive equation for $\bar \tau_s$ and $\bar \tau_\phi=0$ and all geometric relations adopted to the pseudo-surface, the modified shape equations for the wrinkled part are established.

\subsection{Numerical Integration with Automatic Switching between the Shape Equations in Wrinkled Regions}
The shape equations are integrated numerically using a shooting method with $\tau_s(0)$ as the free shooting parameter, which is adjusted until the boundary condition $r(L_0) = a/2$ at the far end is satisfied.

The integration starts at the apex, using the usual shape equations (\ref{eq:shape_eqns}). In each integration step, the wrinkling condition (\ref{eq:wrinkling_cond}) is checked. When $\lambda_\phi$ falls below this threshold, {at $s_0 = s_A$}, the integration is stopped. From this point on, the wrinkled shape equations are integrated, using continuity conditions for all functions as starting conditions. The integration goes on until {the point $s_0 = s_B$, where} the wrinkling condition is not met any more, i.e.\ {where}
\begin{equation}
 \frac{\bar r}{r_0} > 1- \gamma \frac{1-\nu_{2D}^2}{Y_{2D}} \lambda_s - \nu_{2D} (\lambda_s - 1).
\end{equation}
Then we switch back to the original shape equations (\ref{eq:shape_eqns}), again using continuity conditions for all functions. This last part should run up to the end $s_0 = L_0$, where the boundary deviation $r(L_0) - a/2$ can be calculated. The initial guess of $\tau_s(0)$ at the very beginning of the integration is then adjusted, and after some iterations the boundary deviation should become close to zero. {The wrinkle length $L_w$, necessary for the wrinkling analysis, can be obtained as $L_w = s_B - s_A$.}

In some cases, especially for high values of $Y_{2D}$, a simple shooting method will fail to converge. It turned out that these cases are reliably handled by a multiple shooting method \cite{stoer}.

\section{Wavelength of the Wrinkles}
We investigate the wrinkling of a surface that is curved in two directions (see Fig.\ \ref{fig:wrinkling} b) with two different {average} curvatures $\kappa_s = 1/R_s$ and $\kappa_\phi = 1/R_\phi$ which are constant within the wrinkling region. The region is parametrized by the arc lengths $s$ and $t = \phi R_\phi$. Upon wrinkling, the normal displacement of the surface leads to stretching in both directions, and to bending predominantly in $t$ direction because the wavelength $\Lambda$ in $t$-direction is much smaller than in $s$-direction, where we assume only one half sine period. By balancing these main contributions to the deformation energy, we can determine the wrinkling wavelength and critical compressive stress.

The length change of a curved fibre which is displaced sinusoidally in normal direction can be calculated from a parametrization
\begin{equation}
 \vec r(s) = [R+C \sin ks] \zweiervec{\cos s/R}{\sin s/R}
\end{equation}
up to quadratic order in the wrinkle amplitude $C$ as
\begin{align} \label{eq:strain}
 \varepsilon &= \frac{|\diff \vec r|}{\diff s} - 1 \\
 &= \frac{C}{R}\sin ks + \frac{1}{2}\, C^2 k^2 \cos^2 ks + \mathcal{O}(C^3).\nonumber
\end{align}
For the stretch energy in meridional direction, we assume that the wrinkles have length $L_w$ and hence wave vector $k=\pi/L_w$ in $s$-direction. The amplitude of the wrinkles depends on the position along the circumferential direction by $C(t) = C_0 \sin {2\pi t}/{\Lambda}$, and we take $R=R_s$ in (\ref{eq:strain}) to obtain the strain
\begin{equation}
\varepsilon_s(s, t) = \frac{C(t)}{R_s} \sin \frac{\pi s}{L_w} + \frac{\pi^2 C(t)^2}{2 L_w^2}   \cos^2 \frac{\pi s}{L_w}.
\end{equation} 
Upon wrinkling, this strain has to work against the tension $\tau_s$, resulting in a deformation energy
\begin{equation}
 W_s = \int \tau_s \varepsilon_s \, \diff s \, \diff t 
 = \pi^3 C_0^2 R_\phi \tau_s  / 4 L_w 
\end{equation}
where the tension $\tau_s$ was assumed to be constant on the whole integration domain $s \in [0,\,L_w]$ and $t \in [0,\,2\pi R_\phi)$ and the $t$-range is a multiple of $\Lambda$ so that the $t$ integration is performed over full sine periods.

The stretch energy in circumferential direction can be calculated analogously, but with wavevector $k = 2\pi / \Lambda$, $R=R_\phi$ and amplitude $C(s) = C_0 \sin(\pi s / L_w)$ used in (\ref{eq:strain}), and reads
\begin{equation}
 W_\phi = \int \tau_\phi \varepsilon_\phi \, \diff s \, \diff t 
 = \pi^3 C_0^2 L_w R_\phi \tau_\phi / \Lambda^2.
\end{equation}
Since $\tau_\phi<0$, this contribution is negative, i.e.\ it is the energy gain which drives the wrinkling.

The bending is strongest in $\phi$-direction, and its energy cost depends on the curvature change of a circumferential fibre. For a curve given in polar coordinates, $r(\phi) = R_\phi + C \sin {2\pi \phi R_\phi}/{\Lambda}$, the curvature can be approximated to first order in the amplitude $C$ as
\begin{align}
 \kappa_\phi &\approx \frac{1}{r(\phi)} - \frac{r''(\phi)}{r(\phi)^2} \\
 &\approx \frac{1}{R_\phi} + C \left( -\frac{1}{R_\phi^2} + \frac{4\pi^2}{\Lambda^2} \right) \sin \frac{2\pi t}{\Lambda}.
\end{align} 
Considering that the wrinkle amplitude depends on the position along the meridional direction via $C(s) = C_0 \sin \pi s/L_w$, the bending energy reads
\begin{align}
 W_B &= \int \diff s \, \diff t \left\{ \frac{1}{2} \, E_B \left( \kappa_\phi - \frac{1}{R_\phi} \right)^2 \right\} \\
 &= \frac{1}{4} \pi E_B C_0^2 L_w R_\phi \left( \frac{4\pi^2}{\Lambda^2} - \frac{1}{R_\phi} \right)^2.
\end{align}

For the wrinkled state to become preferable to the unwrinkled state, the total deformation energy $W_s + W_\phi + W_B$ must be negative,
\begin{equation}
 \tau_s \frac{\pi^2}{L_w^2} + \tau_\phi \frac{4 \pi^2}{\Lambda^2} + E_B  \left( \frac{4 \pi^2}{\Lambda^2} - \frac{1}{R_\phi^2} \right)^2 < 0.
\end{equation} 
This condition is equivalent to eq.\ (\ref{eq:tau_phi}) in the main text, repeated here for convenience:
\begin{equation} \label{eq:tau_phi_lambda}
 \tau_\phi(\Lambda) = -\tau_s \frac{\Lambda^2}{4 L_w^2} - E_B \frac{\Lambda^2}{4 \pi^2}\left( \frac{4 \pi^2}{\Lambda^2} - \frac{1}{R_\phi^2} \right)^2.
\end{equation}
The wrinkling will first occur with a wavelength that renders $|\tau_\phi(\Lambda)|$ minimal, which is
\begin{equation}
 \Lambda_c = \left( \frac{16 \pi^2 E_B L_w^2}{\tau_s + E_B L_w^2 / \pi^2 R_\phi^4} \right)^{1/4}.
\end{equation}
Solving this equation for $E_B$ yields
\begin{equation} \label{eq:EB}
 E_B = \frac{\tau_s \Lambda^4}{16 \pi^2 L_w^2 \left( 1- \Lambda^4/16 \pi^4 R_\phi^4 \right)},
\end{equation} 
which can be used to determine the bending modulus from measurements of the wrinkle wavelength.

If the wrinkle wavelength is much smaller than the radius of curvature, $\Lambda \ll R_\phi$, the term $1/R_\phi^2$ in (\ref{eq:tau_phi_lambda}) can be neglected and the resulting critical wavelength is exactly the result of Cerda and Mahadevan \cite{Cerda2003},
\begin{equation}
 \Lambda_c = \left( \frac{16 \pi^2 E_B L_w^2}{\tau_s} \right)^{1/4}.
\end{equation}
Note that the small ratio $\Lambda / R_\phi$ enters the formula for $E_B$ (\ref{eq:EB}) in the fourth power, so that the influence of the initial curvature of the membrane has only little influence on the wavelength analysis, and can therefore be neglected.

For the shape equations in the wrinkled region, this compressive circumferential stress is set to zero. That can be justified by considering its order of magnitude: The compressive stress for the critical wavelength reads (in the limit $\Lambda \ll R_\phi$, for simplicity)
\begin{equation}
 \tau_{\phi,c} = \tau_{\phi}(\Lambda_c) = - \sqrt{\frac{4 \pi^2 E_B \tau_s}{L_w^2}}.
\end{equation}
Estimating $E_B \sim E_{2D} H_0^2$ by the relation from classical shell theory and $\tau_s \sim \gamma$ leads to a dimensionless parameter 
\begin{equation}
 \frac{|\tau_{\phi,c}|}{\gamma} \sim \sqrt{\frac{E_{2D}}{\gamma}} \cdot  \frac{H_0}{L_w}.
\end{equation}
Whereas the membrane thickness $H_0$ is of the order of micro- to nanometers, the wrinkle length is around one millimeter. Thus the nondimensionalized critical compression is only of the order $10^{-6}$ to $10^{-3}$.

\bibliography{literature}

\end{document}